\documentclass[12pt]{article}

\usepackage{scicite}

\usepackage{times}
\usepackage{xcolor}
\usepackage{ulem}
\usepackage{graphics,graphicx}
\usepackage{amssymb}
\usepackage{comment}

\usepackage{enumitem,kantlipsum}

\newcommand{\gtt}{\hbox{G323.7$-$1.0}}

\newcommand{\Jftf}{\hbox{HESS\,J1534$-$571}}
\newcommand{\Mftf}{\hbox{MAXI\,J1535$-$571}}

\newcommand{\nat}{\hbox{Nature}}
\newcommand{\apj}{\hbox{The Astrophysical Journal}}
\newcommand{\aap}{\hbox{Astronomy \& Astrophysics}}
\newcommand{\mnras}{\hbox{Monthly Notices of the Royal Astronomical Society}}
\newcommand{\apjl}{\hbox{The Astrophysical Journal Letters}}
\newcommand{\apjs}{\hbox{The Astrophysical Journal Supplement}}
\newcommand{\aj}{\hbox{The Astronomical Journal}}
\newcommand{\araa}{\hbox{Annual Review of Astronomy \& Astrophysics}}

\title{{A supernova remnant associated with a nascent black hole low-mass X-ray binary}} 

\author
{Nigel I. Maxted,$^{1\ast}$ Ashley J. Ruiter,$^{1}$ Krzysztof Belczynski,$^{2}$\\ Ivo R. Seitenzahl$^{1}$ and Roland M. Crocker$^{3}$\\
\\
\normalsize{$^{1}$School of Science, The University of New South Wales, Australian Defence Force Academy,}\\
\normalsize{Canberra, ACT 2600, Australia}\\
\normalsize{$^{2}$Nicolaus Copernicus Astronomical Center, Polish Academy of Sciences, Bartycka 18, 00-716 }\\
\normalsize{Warsaw, Poland}\\
\normalsize{$^{3}$Research School of Astronomy and Astrophysics, Australian National University, Canberra}\\
\normalsize{ACT 2611, Australia}\\
}

\date{}

\begin{document} 


\baselineskip12pt


\maketitle



ABSTRACT: Energy released when the core of a high-mass star collapses into a black hole often powers an explosion that creates a supernova remnant. Black holes have limited windows of observability, and consequently are rarely identified in association with supernova remnants. Analysing multi-messenger data, we show that \Mftf\ is the black hole produced in the stellar explosion that gave rise to the supernova remnant \gtt , making it the first case of an association between a black hole low-mass X-ray binary and a supernova remnant. Given this connection, we can infer from our modeling that the progenitor system was a close binary whose primary star had an initial mass of $\sim$23-35 solar masses with a companion star about 10 times less massive.\\



The lives of certain stars with birth masses greater than $\sim$20-30\,M$_{\odot}$ 
end in violent events that feature the implosion of the stellar core into a black hole as, simultaneously, the outer stellar layers are explosively ejected into a shell of shocked material to form a `supernova remnant' (SNR). While SNRs aged hundreds to tens of thousands of years are often readily observable in our Galaxy, the detection of stellar-mass ($\sim$3-50$\times$ the mass of the sun) black holes (BHs) is contingent upon
interaction of the BH with a partner star in orbit. Such a BH binary companion could be a neutron star or another BH, as in the case of gravitational wave signatures of BH merger events seen by the LIGO and Virgo experiments \cite{Abbott_2019_BHGWs}, a star with a measurable radial velocity \cite{Thompson:2019}, or a companion `donor' star that facilitates the creation of an X-ray Binary (XRB) system through accretion of stellar material onto the BH \cite{Remillard_2006_XRB}. The detection of the SNR counterpart for a BH would significantly constrain the input physics relevant to BH formation, SNR evolution and the history of the system in question. 
An enduring claim for a connection between a BH and a SNR has previously only been made for one system - the eclipsing high-mass X-ray binary (HMXB) SS\,433 \cite{Ryle:1978_w50}, which likely 
consists of a 10-20\,M$_{\odot}$ BH and a more massive supergiant star \cite{Margon:1984}. W50 is the radio nebula widely attributed to being the SNR associated with SS\,433, though its origin is still debated \cite{Farnes:2017}. Here, we make the case for the first association of a BH and a SNR for a low-mass X-ray binary (LMXB), which is fed via Roche lobe overflow through an accretion disk rather than winds; the latter is typically the case for HMXBs \cite{Casares:2017}.

As we now explain, the strong BH candidate \Mftf\ is likely a BH that originated from the same explosion that created the supernova remnant, \gtt , otherwise known as \Jftf when referring to the corresponding shell-like gamma-ray emission \cite{Abramowski:2017newshells}. Both the SNR and BH are at a consistent distance and the BH is located towards the central region of the SNR. According to our coincidence modelling, this is unlikely a coincidence and instead indicates the objects are related.

\begin{sloppypar}
In order to characterise the chance likelihood of discovering a black hole such as \Mftf\ coincident to a SNR such as \gtt , we performed simulations that randomly generated 10$^5$ galaxies with SNR and BH positions sampled from a Milky Way-like distribution. We generated Kernel Density Estimations (KDE) from the positions of known BHs and SNRs according to Milky Way catalogs \cite{Corral-Santana:2016,Green:2019_snrcat}. 60 observable BHs and 300 observable SNRs were assumed per galaxy (similar to the Milky Way), and for simplicity, an average radius value was used for elliptical Milky Way SNRs in the KDE generation process. 
\end{sloppypar}

We defined a chance coincidence leading to a false association as a BH position being within one third of the SNR radius of a SNR position. Using this criteria, an observed stellar-mass BH has a $\sim$10$^{-3}$\% chance of being incorrectly associated with an observed SNR in a given Milky Way-like galaxy, while a a false association will only occur in $<$0.1\% of Milky Way-like galaxies. We can thus assert {\it from positional coincidence alone}, that \Mftf\ and \gtt\ 
very likely originated from the same supernova event. 
The false association probability derived from the position can be considered an upper limit because it does not account for the probability of coincidental overlap for the 3rd dimension - distance. Furthermore, it does not quantify the added confidence gained from multi-wavelength data of a corresponding cavity in atomic gas, and a consistent evolutionary path description for the binary star (as described later).

As derived from independent investigations, the estimated distances of the SNR and BH candidate are consistent with one another, thus strongly reinforcing a scenario of physical association rather than chance-alignment. Analysis of hydrogen absorption lines in the radio emission spectrum of \Mftf\ constrain the probable distance to 4.1$^{+0.6} _{-0.5}$\,kpc \cite{Chauhan:2019}. This is consistent with the proposed wind-blown cavity of the \gtt\ supernova progenitor star, 
which has a distance of $\sim$3.5\,kpc, as estimated from Galactic rotation modelling of the line of sight velocity of associated spectral lines \cite{Maxted:2018_HESSJ1534}.
Independent of the \Mftf\ discovery, the existence of the cavity (Figure\,1) favours a core-collapse supernova explosion model for the SNR, like that required to create a BH. The cavity is also consistent with SNR distance estimates from radio brightness \cite{Maxted:2018_HESSJ1534}, X-ray absorption \cite{Saji:2018} and gamma-ray emission modeling \cite{Araya:2017}. Given the strong case for a relation between the SNR and BH, 
we used the SNR shock evolution modelling age estimate of 8-24\,kyr \cite{Maxted:2018_HESSJ1534} as a guide to model the history of this unique association. 

\begin{figure}
\includegraphics[width=\columnwidth]{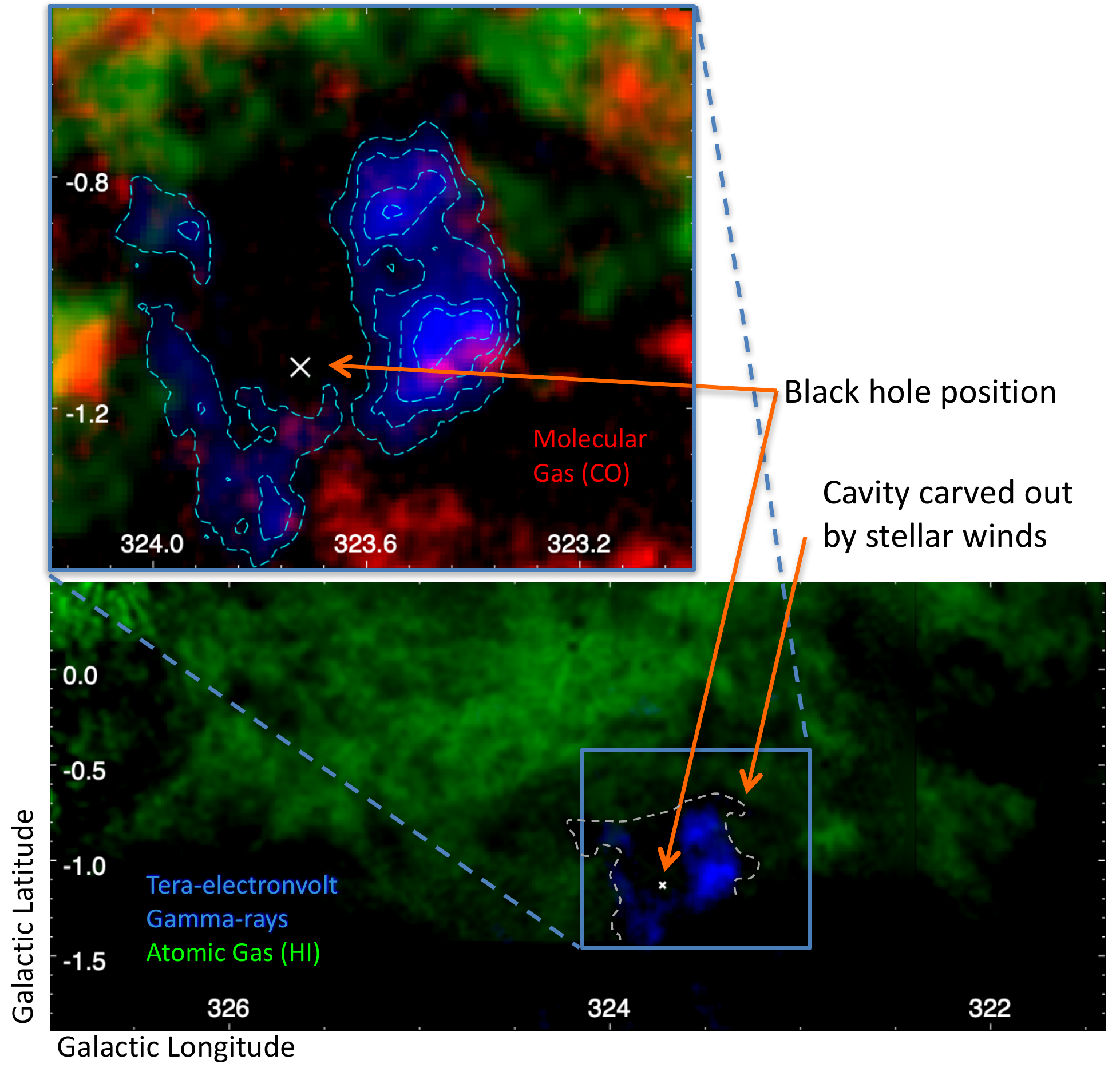}
\caption{
The supernova remnant within its progenitor stellar wind-blown gas cavity. The top image is a compilation of HESS TeV gamma-rays (blue) that shows the shell structure of the supernova remnant HESS\,J1534$-$571 (G323.7$-$1.0) \cite{Abramowski:2017newshells}, Southern Galactic Plane Survey 21 cm HI emission (green) that shows a cavity in atomic gas and Mopra CO(1-0) emission (red), which is a proxy for molecular gas clouds \cite{Maxted:2018_HESSJ1534,McClure:2005}. The bottom image highlights the atomic cavity and the large scale structure of the HI emission. The position of black hole, MAXI J1535-571, is indicated in each image. }
\end{figure}

A particularly exciting implication of the association we have uncovered is the scenario required  such that it be discoverable. To explain its recent X-ray activity, the BH must have formed from the supernova explosion of a progenitor orbited by a pre-existing companion. This companion needed to stay orbitally-bound to the BH post-supernova, then quickly form an accretion disk, through the process of Roche-lobe overflow, while the SNR shell remained visible to the observer.

\begin{sloppypar}
We used the binary population synthesis code \textsl{StarTrack} {\cite{Belczynski:2008_StarTrack}}
to explore binary star evolution scenarios consistent with {the} \Mftf\ - \gtt\ connection.
{P}rimary and secondary star initial masses {are} the main inputs into our models. The \gtt\ SNR progenitor star needed to be capable of forming a \Mftf\ BH mass between $\sim$5 and $\sim$11\,M$_{\odot}$, consistent with estimates made with the wealth of data collected during the active phase of the X-ray flaring of \Mftf , and before the XRB entered into a quiescent phase in October 2017. We identified three main binary evolution scenarios that can create a BH of mass $\sim$5-11\,M$_{\odot}$ while retaining a small donor companion star of low mass (e.g., unseen) that is in a close-orbit configuration that facilitates mass-transfer within a short timescale post-SN. This is {essential for}  X-ray flaring {to occur} while the SNR is still present, as we observe for \Mftf. 
\end{sloppypar}

\begin{sloppypar}
Three plausible scenarios that successfully result in systems consistent with \gtt\ and \Mftf\ have been found in our models. The key evolutionary phases are presented in Figure\,2. For brevity we outline only one evolutionary path as an example here. Observations of the XRB and SNR can be simultaneously described by a model that begins with a primary star of mass $M_{ZAMS} =$24.4\,M$_{\odot}$ and a 2.6\,M$_{\odot}$ secondary companion in a $\sim$3200\,R$_{\odot}$ low-eccentricity ($e\sim$0.1) orbit. Wind mass loss removes ${\sim} 5$ M$_{\odot}$ from the primary as it evolves, eventually overfilling its Roche lobe ${\sim} 8.8$ \,Myr after binary formation, when it has become an asymptotic giant branch star. This results in a common envelope (CE) event between the 19.0 M$_{\odot}$ asymptotic giant branch primary and the main sequence companion. The primary star collapses to form a black hole of mass $5.6$ M$_{\odot}$ soon after the CE is initiated,
which should lead to supernova ejecta interacting with the primary's lost envelope. If we include the mass lost during the CE phase when calculating the ejected mass, which is a good assumption given that the CE and SN occur within a very short time\footnote{The time resolution of analytic formulae used in our evolutionary calculations (see e.g. \cite{Hurley:2000}) does not allow for precise time estimate between CE and SN, but they happen in two consecutive time steps that may vary from several years to ${\sim} 1400$ years for this particular binary system.}, we estimate a total mass of ejecta M$_{ej}\sim$13.4\,M$_{\odot}$.\end{sloppypar}

After the common envelope, the binary is on a circular orbit with a separation of 25 R$_{\odot}$. The supernova imparts an asymmetric natal kick on the BH, and post-SN there is some fallback onto the black hole and the separation decreases to 13.7 R$_{\odot}$ (the formation of the BH imparts a high eccentricity of $e=0.8$ on the orbit). Tidal interactions (e.g. a tidal bulge is raised on the companion) result in a new, smaller separation of 3\,R$_{\odot}$, equivalent to the separation at periastron passage \cite{Belczynski:2008_StarTrack}, as the binary quickly circularises. We note that even in the extreme case where tidal interactions are neglected, which would result in a slightly larger separation, $a_{\rm new}=a_{\rm old}(1-e^2)$, the overall results do not change.  
Mass transfer from the companion onto the BH starts via Roche-lobe overflow immediately post-supernova. Plausible binary evolution scenarios of primary stars spanning initial masses ($M_{ZAMS}$) as low as 23 \,M$_{\odot}$, and as high as ${\sim}$35\,M$_{\odot}$ can lead to the formation of \Mftf\ and \gtt . While primary masses outside of this mass range seem unlikely for progenitors of LMXBs, they cannot be confidently ruled out until a more comprehensive exploration of the parameter space has been carried out. 
We did, however, calculate the likelihood that such a binary configuration exists in our Milky Way from the standpoint of binary evolution: In this case a black hole that begins accreting via Roche-lobe overflow from a stellar companion within 24,000 yr of supernova explosion, still visible within its own remnant. Taking a maximum visible SNR age of 100,000 yr, a constant star formation rate for the Galaxy with a total mass in stars of $6.4 \times 10^{10}$ M$_{\odot}$, and a 70\% binary fraction, we estimate using StarTrack there are ${\sim} 2-3$ such systems currently residing in the MW Galaxy, which is consistent with our findings.  

\begin{figure}
\includegraphics[width=\columnwidth]{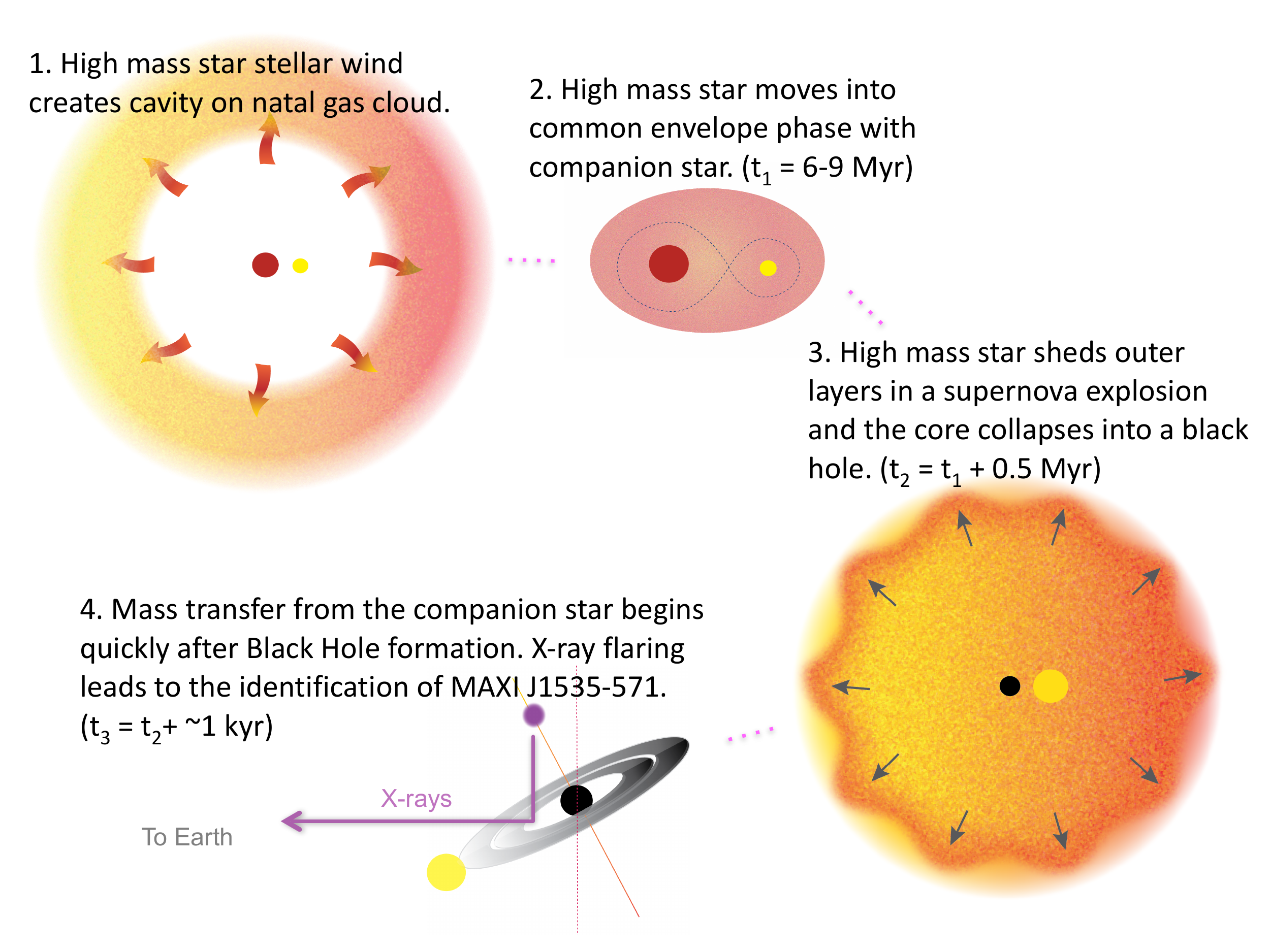}
\caption{A schematic of a plausible evolutionary pathway leading to the formation of the black hole, MAXI\,J1535$-$571, and the associated and supernova remnant.}
\end{figure}

\begin{sloppypar}
For a star of $M_{ZAMS} = $23-35\,M$_{\odot}$, such as we propose for the progenitor for the \Mftf\ BH, the stellar wind pressure would become balanced by an average ISM inter-clump pressure for a wind-blown bubble of diameter of {43$\pm$4 to 67$\pm$4} according to the relationship presented by \cite{Chen:2013_BubbleSize}. 
This range is a match to the dimensions of the atomic bubble proposed to be associated with \gtt\ (see Figure\,1, \cite{Maxted:2018_HESSJ1534}). 
At a distance of $d_{\tiny{\textrm{kpc}}}$, the SNR cavity of angular size 51$^{\prime}\times$38$^{\prime}$ would have a linear size of (44-59)($\frac{d_{\tiny{\textrm{kpc}}}}{4\,\tiny{\textrm{kpc}} }$)\,pc, thus matching the $\sim$43-67\,pc prediction from stellar pressure models \cite{Chen:2013_BubbleSize} for a 4\,kpc distance.
This further strengthens the binary evolution scenarios proposed in our study, while reinforcing the 4\,kpc distance in a self-consistent way. 
\end{sloppypar}


With regard to observational signatures of the XRB, note that while optical campaigns have uncovered a likely flaring/flickering jet counterpart to the black hole
\cite{Scaringi:2017ATel,Baglio:2018}, a candidate companion star has not been detected.
In fact, this non-detection is entirely consistent with our scenario:
Our binary population synthesis modeling points to a secondary star of mass 
${\sim}$2.6-4.5\,M$_{\odot}$, which will have an absolute magnitude of ${\sim} -1$, assuming the star is late B-type. 
Approximately 13-15 magnitudes \cite{Schlafly:2011,Schlegel:1998} of optical extinction is expected from the $\gtrsim$ 2$\pm$1$\times$10$^{22}$\,cm$^{-2}$ of foreground proton column density \cite{Chauhan:2019}. A further increase of $5\log_{10}{(\frac{10}{d_{pc}})}$ 
to the value of the apparent magnitude is due to a distance, $d_{pc}$, of 4$\times$10$^3$\,pc. It follows that the expected total apparent magnitude of the companion star is $\sim$25-27 - 
undetectable given the current depth of observations.

The binary star configurations that we have uncovered as a good match to explain the connection between \Mftf\ and {\gtt} are also good examples of X-ray binary evolution. These binary systems - which could be labeled as `intermediate-mass XRBs' (since their donors are $\sim$a few M$_{\odot}$) evolve through stable Roche-lobe overflow which removes mass from the donors until they become low-mass stars ($< 1$ M$_{\odot}$) and the binaries become LMXB transients similar to those observed in our Galaxy \cite{Ritter:2003}. 

 Our scenarios require SN natal kicks (based on \cite{Hobbs:2005} and accounting for the amount of fallback that is expected during BH formation \cite{fryer:2012}), which decrease the orbital separation and lead to Roche-lobe overflow (RLOF) soon after BH formation. The absolute kick velocities that our three evolutionary example BHs receive ($\sim$150 -- 200 km/s) are on the order of those of observed BHs XTE J1118+480 and XTE J1819-254 \cite{Belczynski:2016_BHkicks}. BH natal kicks are rather uncertain, and low to moderately-high BH natal kicks are consistent with observations \cite{Repetto:2017_BHs,Mandel:2016_kicks}. 
As discussed previously, we do not calculate the timescale for tidal circularisation in cases when a star overfills its Roche lobe. In such cases, we assume that circularisation is instantaneous and we circularise  the binary to its periastron distance (with circularisation of the orbit via tidal interactions). 
 Even in the case where circularisation would take a long time (unlikely), the episodic/eccentric RLOF mass transfer would still occur and the accretion disk around the BH would be fed with mass, leading to X-ray activity.
 Though our treatment for circularisation is arguably the preferred way to treat close, eccentric binary stars \cite{Zahn:2005_tides}, we reiterate that our main conclusions remain unaffected even if we adopt a different treatment for orbital evolution: if we assume explicitly in our simulations that tidal-spin interactions play no role in the orbital circularisation, we still form a RLOF mass transfer binary soon after BH formation that fits the description of the black hole MAXI\,J1535$-$571 and the associated supernova remnant \gtt. 
 
Finally, we note that we investigated the potential of the future gravitational wave detector, the Laser Interferometer Space Antenna (LISA), to observe this binary today, and found that with a strain amplitude of $\sim$10$^{-22}$ and a gravitational wave frequency $\sim$10$^{-5}$\,Hz, the binary would be outside of LISA's currently-planned sensitivity range.


\begin{thebibliography}{10}

\bibitem{Abbott_2019_BHGWs}
B.~P. {Abbott}, {\it et~al.\/}, {\it \apjl\/} {\bf 882}, L24 (2019).

\bibitem{Thompson:2019}
T.~A. {Thompson}, {\it et~al.\/}, {\it Science\/} {\bf 366}, 637 (2019).

\bibitem{Remillard_2006_XRB}
R.~A. Remillard, J.~E. McClintock, {\it Annual Review of Astronomy and
  Astrophysics\/} {\bf 44}, 49 (2006).

\bibitem{Ryle:1978_w50}
M.~{Ryle}, J.~L. {Caswell}, G.~{Hine}, J.~{Shakeshaft}, {\it \nat\/} {\bf 276},
  571 (1978).

\bibitem{Margon:1984}
B.~{Margon}, {\it \araa\/} {\bf 22}, 507 (1984).

\bibitem{Farnes:2017}
J.~S. {Farnes}, {\it et~al.\/}, {\it \mnras\/} {\bf 467}, 4777 (2017).

\bibitem{Casares:2017}
J.~{Casares}, P.~G. {Jonker}, G.~{Israelian}, {\it {X-Ray Binaries}\/} (2017),
  p. 1499.

\bibitem{Abramowski:2017newshells}
{H.E.S.S.~Collaboration}, {\it et~al.\/}, {\it \aap\/} {\bf 612}, A8 (2018).

\bibitem{Corral-Santana:2016}
J.~M. {Corral-Santana}, {\it et~al.\/}, {\it \aap\/} {\bf 587}, A61 (2016).

\bibitem{Green:2019_snrcat}
D.~A. {Green}, {\it Journal of Astrophysics and Astronomy\/} {\bf 40}, 36
  (2019).

\bibitem{Chauhan:2019}
J.~{Chauhan}, {\it et~al.\/}, {\it \mnras\/} {\bf 488}, L129 (2019).

\bibitem{Maxted:2018_HESSJ1534}
N.~I. {Maxted}, {\it et~al.\/}, {\it \mnras\/} {\bf 480}, 134 (2018).

\bibitem{McClure:2005}
N.~M. {McClure-Griffiths}, {\it et~al.\/}, {\it \apjs\/} {\bf 158}, 178 (2005).

\bibitem{Saji:2018}
S.~{Saji}, {\it et~al.\/}, {\it Publications of the Astronomical Journal of
  Japan\/} {\bf 70}, 23 (2018).

\bibitem{Araya:2017}
M.~{Araya}, {\it \apj\/} {\bf 843}, 12 (2017).

\bibitem{Belczynski:2008_StarTrack}
K.~Belczynski, {\it et~al.\/}, {\it The Astrophysical Journal Supplement
  Series\/} {\bf 174}, 223 (2008).

\bibitem{Hurley:2000}
J.~R. {Hurley}, O.~R. {Pols}, C.~A. {Tout}, {\it \mnras\/} {\bf 315}, 543
  (2000).

\bibitem{Chen:2013_BubbleSize}
Y.~{Chen}, P.~{Zhou}, Y.-H. {Chu}, {\it \apjl\/} {\bf 769}, L16 (2013).

\bibitem{Scaringi:2017ATel}
S.~{Scaringi}, {ASTR211 Students}, {\it The Astronomer's Telegram\/} {\bf
  10702} (2017).

\bibitem{Baglio:2018}
M.~C. {Baglio}, {\it et~al.\/}, {\it \apj\/} {\bf 867}, 114 (2018).

\bibitem{Schlafly:2011}
E.~F. {Schlafly}, D.~P. {Finkbeiner}, {\it \apj\/} {\bf 737}, 103 (2011).

\bibitem{Schlegel:1998}
D.~J. {Schlegel}, D.~P. {Finkbeiner}, M.~{Davis}, {\it \apj\/} {\bf 500}, 525
  (1998).

\bibitem{Ritter:2003}
H.~{Ritter}, U.~{Kolb}, {\it \aap\/} {\bf 404}, 301 (2003).

\bibitem{Hobbs:2005}
G.~{Hobbs}, D.~R. {Lorimer}, A.~G. {Lyne}, M.~{Kramer},\\ {\it \mnras\/} {\bf 360}, 974 (2005).

\bibitem{fryer:2012}
C.~L. {Fryer}, {\it et~al.\/}, {\it \apj\/} {\bf 749}, 91 (2012).


\bibitem{Belczynski:2016_BHkicks}
K.~{Belczynski}, {\it et~al.\/}, {\it \apj\/} {\bf 819}, 108 (2016).

\bibitem{Repetto:2017_BHs}
S.~{Repetto}, A.~P. {Igoshev}, G. {Nelemans}, {\it \mnras} {\bf 467}, 298 (2017). 

\bibitem{Mandel:2016_kicks}
I.~Mandel, {\it \mnras} {\bf 456}, 578 (2016).

\bibitem{Zahn:2005_tides}
J.-P.~{Zahn}, {\it Astronomical Society of the Pacific Conference Series} {\bf 333}, 4 (2005). 

\end{thebibliography}

\clearpage

\section*{Acknowledgments}
\begin{sloppypar}
{
We acknowledge Australian Research Council support through the Linkage, Infrastructure, Equipment and Facilities project LE160100094 and the Future Fellowship projects FT170100243 and FT160100028. This research was undertaken with the assistance of resources and services from the National Computational Infrastructure (NCI), which is supported by the Australian Government, through the National Computational Merit Allocation Scheme and the UNSW HPC Resource Allocation Scheme. KB acknowledges support from the Polish National Science Center (NCN) grant Maestro (2018/30/A/ST9/00050).
N.I.M. initiated the project, conducted a literature review, carried out the multi-wavelength analysis, conducted positional chance coincidence modelling and created the figures. A.J.R. examined plausible physical interpretations for the association through binary population synthesis simulations and analysis, discovered the theoretical evolutionary configurations that align with the observational data, and guided the project direction. K.B. initiated the binary population synthesis code development, reviewed (and contributed to) the literature on Galactic black holes, and provided binary evolution interpretation. I.R.S. and R.M.C. provided project guidance and consistent feedback about the physics of the association. N.I.M., A.J.R., K.B., I.R.S. and R.M.C. all contributed to the writing of the final manuscript.
}
\end{sloppypar}



\end{document}